\documentclass{appolb}
\usepackage{graphicx}
\usepackage[utf8]{inputenc} 
\usepackage[colorlinks]{hyperref}
\usepackage{amsmath,amssymb}
\usepackage{subcaption}
\usepackage{floatrow}

\begin{document}
\title{Studying particle production in small systems through correlation measurements in ALICE\thanks{Presented at \emph{Excited QCD 2020}, Krynica-Zdrój, Poland, 2-8 February 2020.}%
}
\author{Jonatan Adolfsson
\address{for the ALICE Collaboration}
\address{Lund University, Lund, Sweden}
\\
}
\maketitle
\begin{abstract}
In these proceedings, measurements of angular correlations between hadron pairs in pp collisions obtained by the ALICE experiment at the LHC are presented and compared with phenomenological predictions. Correlations between particles carrying the same and opposite quantum numbers are studied to understand the hadron production mechanism, and the difference between same-sign and opposite-sign correlations is used to probe charge-dependent effects in particle production. Correlation measurements dominated by minijet fragmentation agree well with the models, but other results, in particular correlations between baryons and strange hadrons, are not yet understood.
\end{abstract}

  
\section{Introduction}
In heavy-ion collisions, it is believed that a quark-gluon plasma (QGP) is formed. Two of its key signatures -- both collective in origin -- are flow and strangeness enhancement, which have also been observed in high-multiplicity proton-proton and proton-nucleus collisions \cite{Near-side ridge, Strangeness enhancement}. Due to a lack of other known QGP signatures in these systems, such as jet quenching, and since these systems are expected to be too small to reach thermal equilibrium, these observations are not yet understood.

Several phenomenological models are being developed to try to understand these observations. Two different approaches are being explored. In one approach, QCD inspired models are extended with new features to add collective behaviour to the system. Alternatively, some models form a two-phase state with a dense QGP like core and a dilute corona (core-corona model). The former is used in PYTHIA \cite{PYTHIA8 manual}, where colour reconnection is included in PYTHIA 8.2 and rope hadronisation and string shoving are available in the Angantyr extension \cite{Angantyr}. The latter approach is used in EPOS \cite{EPOS-LHC}.

These models use fundamentally different mechanisms, but neither of them has been unambiguously proven so far. More experimental input is required to be able to distinguish between them, and in particular pinpoint the hadron production mechanism. One such observable is angular correlations between different hadronic species, which can be used both to study quark production early in the collision and hadron production in the later stages. In PYTHIA, hadrons are formed through $q\bar{q}$ pair formation during breakings of colour strings \cite{PYTHIA6 manual}, yielding strong local correlations between hadrons sharing a quark-antiquark pair. On the other hand, if quarks are formed early in the collision and hadron formation happens later, such as in a core-corona model, such correlations should be weak or non-existent.

One can study $\pi-\pi$ or $\rm K-K$ correlations, and $\rm p-p$ or $\Lambda-\Lambda$ correlations, to probe meson and baryon production, respectively. To study strangeness production, one can trigger on a multistrange hadron, such as the $\mathrm{\Xi}$ baryon, and measure its correlation with other strange hadrons, such as kaons.

The correlation function will inevitably be affected by an excess of particles formed in the jet cone (minijet fragmentation), leading to a strong near-side peak. This can be studied through $\pi-\pi$ correlations, since the particle production is dominated by pions. To remove these effects, one can study the \emph{balance function}, which is the difference between correlation functions of equally- and oppositely-charged particle pairs. For a thermally expanding medium, the balance function width is expected to decrease with increasing collision multiplicity due to radial flow \cite[and refs. therein]{Balance functions}.

\section{Method}
Three similar quantities are being measured in the studies presented here: the \emph{correlation function}, the \emph{per-trigger yield}, and the \emph{balance function}. The correlation function is the distribution of particle pairs in relative (pseudo)rapidity ($\Delta \eta$ or $\Delta y$)--azimuthal-angle ($\Delta\varphi$) space, normalised to the number of trigger-associated particle pairs $N_\mathrm{pairs}$:

\begin{equation}
\mathbb{C}(\Delta\eta,\Delta\varphi)=\dfrac{1}{N_{\rm pairs}}\dfrac{\mathrm{d}^2N_{\rm pairs}}{\mathrm{d}\Delta\eta\mathrm{d}\Delta\varphi}.
\label{eq:correlation_function}
\end{equation}
The same distribution, but normalised to the number of triggers $N_\mathrm{trig}$ instead, is called the per-trigger yield:

\begin{equation}
\mathbb{Y}(\Delta y,\Delta\varphi)=\dfrac{1}{N_{\rm trig}}\dfrac{\mathrm{d}^2N_{\rm pairs}}{\mathrm{d}\Delta y\mathrm{d}\Delta\varphi}.
\label{eq:per-trigger_yields}
\end{equation}
Finally, the balance function is defined as

\begin{equation}
\mathbb{B}(\Delta y,\Delta\varphi)=\dfrac{1}{2}\left(\mathbb{Y}_{(+,-)} + \mathbb{Y}_{(-,+)} - \mathbb{Y}_{(+,+)} - \mathbb{Y}_{(-,-)}\right),
\label{eq:balance_functions}
\end{equation}
where the subscript denotes the charge combination.

These quantities are measured in pp collisions ($\sqrt{s}=5.02,7,$ and 13 TeV depending on the analysis) for tracks reconstructed with the ALICE detector \cite{ALICE detector} at the LHC. Pions, kaons, and protons are identified via the specific energy loss in the Time Projection Chamber and the velocity in the Time-Of-Flight subdetectors ($|\eta|<0.8$). The $\Xi$ and $\Lambda$ baryons are reconstructed from their decay products, $\rm\Xi^-\rightarrow\pi^- + \Lambda\rightarrow \pi^-+\pi^-+p$, by making use of their invariant masses and various topological cuts (similar to what is done in Ref. \cite{Strangeness production}). Moreover, for the balance functions, the events are divided into multiplicity classes, where the lowest percentages correspond to the highest multiplicities, which are measured by the V0 subdetector in the forward regions ($-3.7<\eta<-1.7$ and $2.8<\eta<5.1$).

\section{Results}
\subsection{Hadron-hadron correlation functions at 7 TeV}
\label{sec:correlation_results}

\begin{figure}[htb]
\floatbox[{\capbeside\thisfloatsetup{capbesideposition={right,top}, capbesidewidth=0.25\textwidth}}]{figure}[\FBwidth]
{\caption{Projections of same-sign correlations onto $\Delta\varphi$ along with predictions for different versions of PYTHIA and PHOJET, for \textbf{(a)} pions, \textbf{(b)} kaons, \textbf{(c)} protons, and \textbf{(d)} $\rm\Lambda$ baryons \cite{ALICE correlation results}.}\label{fig:projections_like_sign}}
{\includegraphics[width=0.71\textwidth]{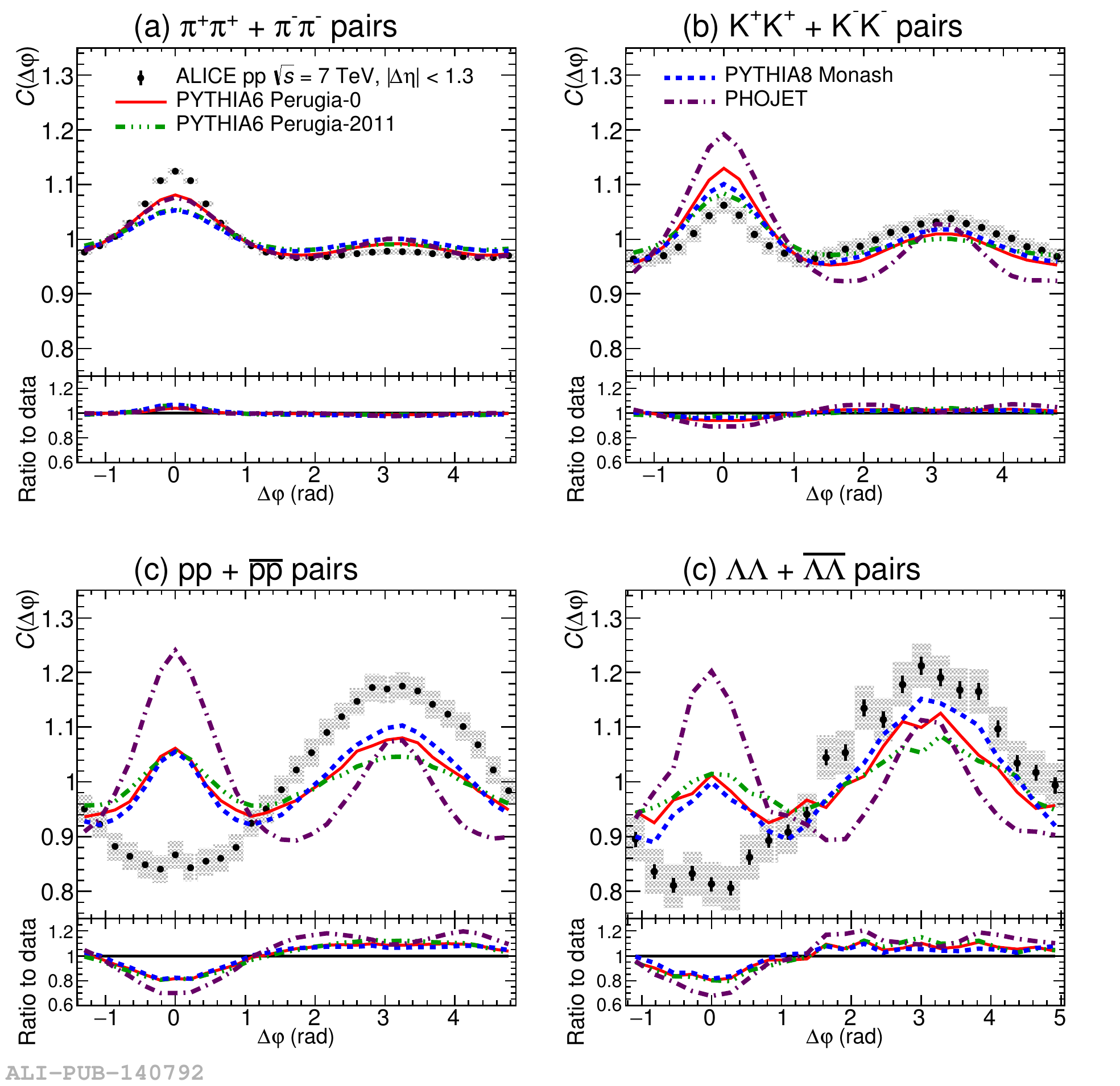}}
\end{figure}

Correlation functions (Eq. \ref{eq:correlation_function}) for $\rm \pi-\pi$, $\rm K-K$, $\rm p-p$, and $\Lambda-\Lambda$ particle pairs are shown in $(\Delta y,\Delta\varphi)$ space in Ref. \cite{ALICE correlation results}. Projections onto $\Delta\varphi$ are shown in Fig. \ref{fig:projections_like_sign} for same-sign particle pairs, along with model predictions from PYTHIA and PHOJET \cite{PHOJET}. In PHOJET, hadronisation is described through Pomeron chain fragmentation. Opposite-sign correlations (shown in Ref. \cite{ALICE correlation results}) and same-sign meson-meson correlations have a strong near-side peak, which is due to minijet fragmentation. This is balanced by an away-side ridge, due to momentum conservation. These features also show up in the models, but PYTHIA describes the correlation function better than PHOJET, favouring hadronisation through $q\bar{q}$ string breakings.

On the other hand, for $\rm p-p$ and $\Lambda-\Lambda$ correlations of particles of the same baryon number, there are near-side anti-correlations, meaning that production of these particles close to each other in phase space is disfavoured. This is not reproduced in either model, meaning that baryon production mechanism is not fully understood. In the string model, two equal baryons cannot be produced close in phase space, so intuitively some near-side suppression is expected, but this is not enough to explain the data.

\subsection{Balance functions at 5.02 TeV}
\begin{figure}[htb]
\centerline{
\begin{subfigure}{0.49\textwidth}
\includegraphics[width=\textwidth]{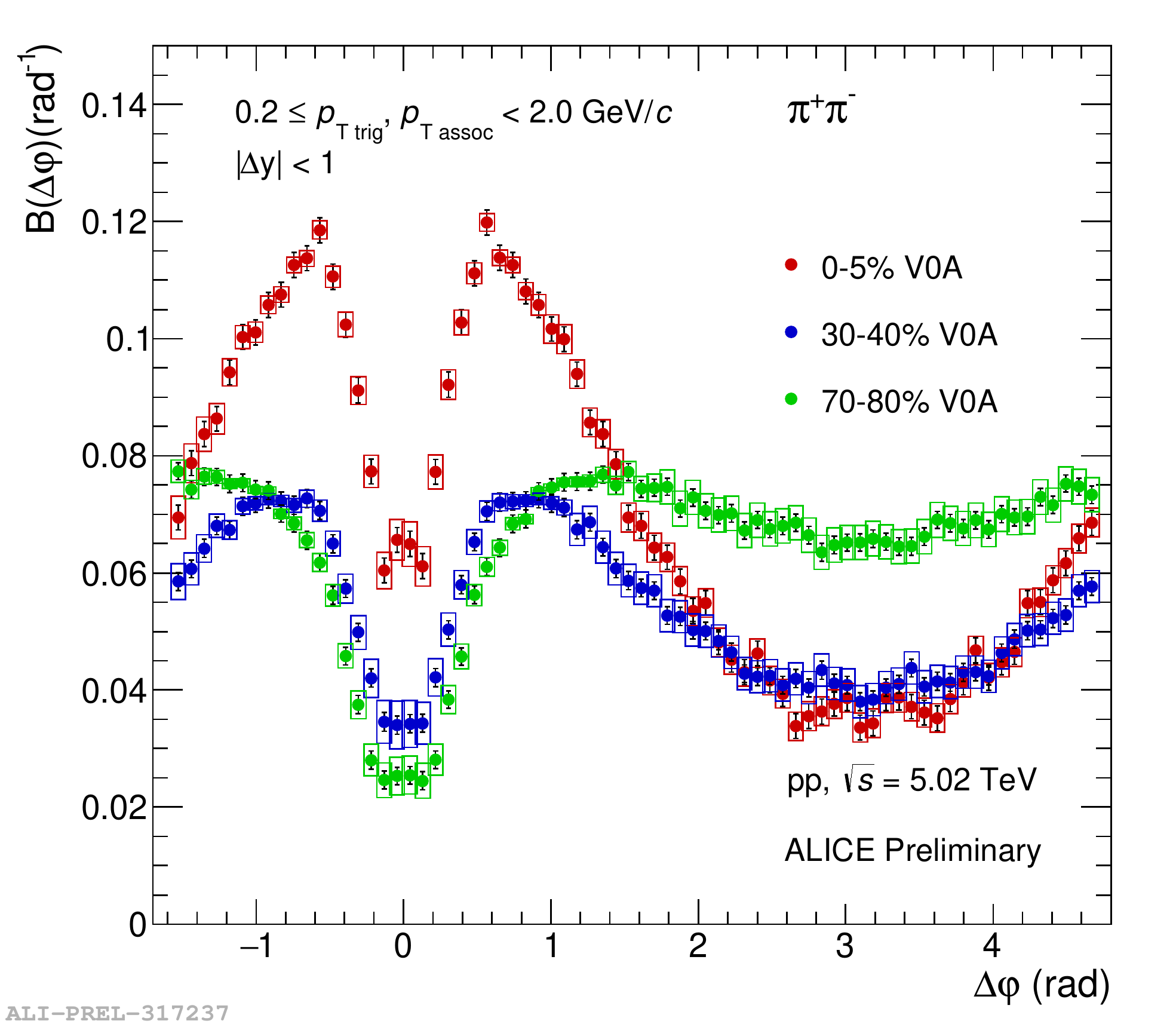}
\caption{}
\end{subfigure}
\begin{subfigure}{0.49\textwidth}
\includegraphics[width=\textwidth]{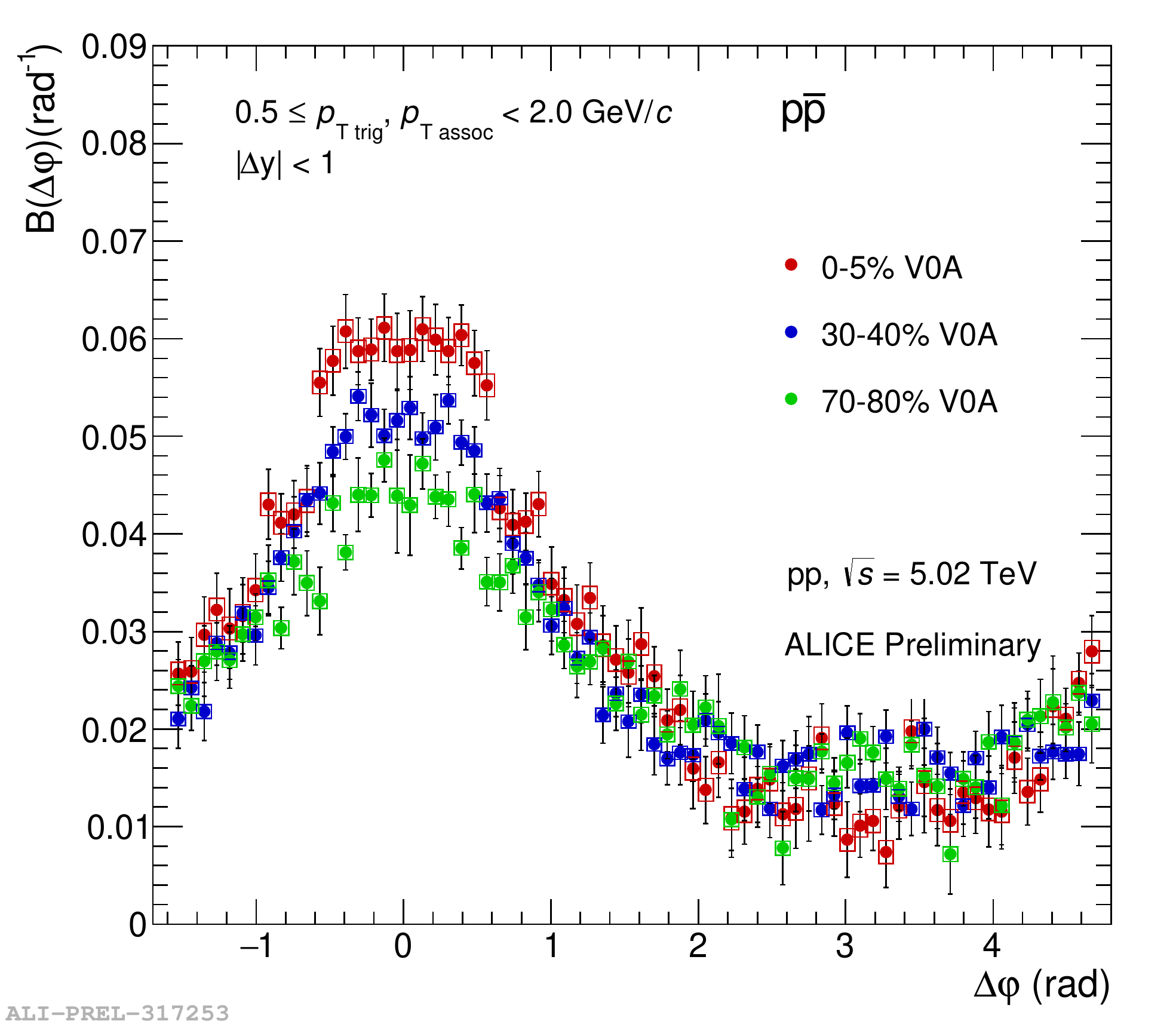}
\caption{}
\end{subfigure}
}
\caption{Balance functions projected onto $\Delta\varphi$, for \textbf{(a)} pions in the momentum range $0.2\leq p_\mathrm{T}<2.0\,\mathrm{GeV}/c$ and \textbf{(b)} protons at $0.5\leq p_\mathrm{T}<2.0\,\mathrm{GeV}/c$, for a few different multiplicity classes.}
\label{fig:balance_functions}
\end{figure}

Balance functions (Eq. \ref{eq:balance_functions}) at low momentum for pions and protons, projected onto $\Delta\varphi$, are shown in Fig. \ref{fig:balance_functions}. These have positive values, in particular on the near side, meaning that opposite-sign correlations are stronger than same-sign, which agrees with the results presented in Sec. \ref{sec:correlation_results}. The dip around $\Delta\varphi=0$ for pions is due to quantum statistics \cite{Balance functions}.

\begin{figure}[htb]
\centerline{
\begin{subfigure}{0.499\textwidth}
\includegraphics[width=\textwidth]{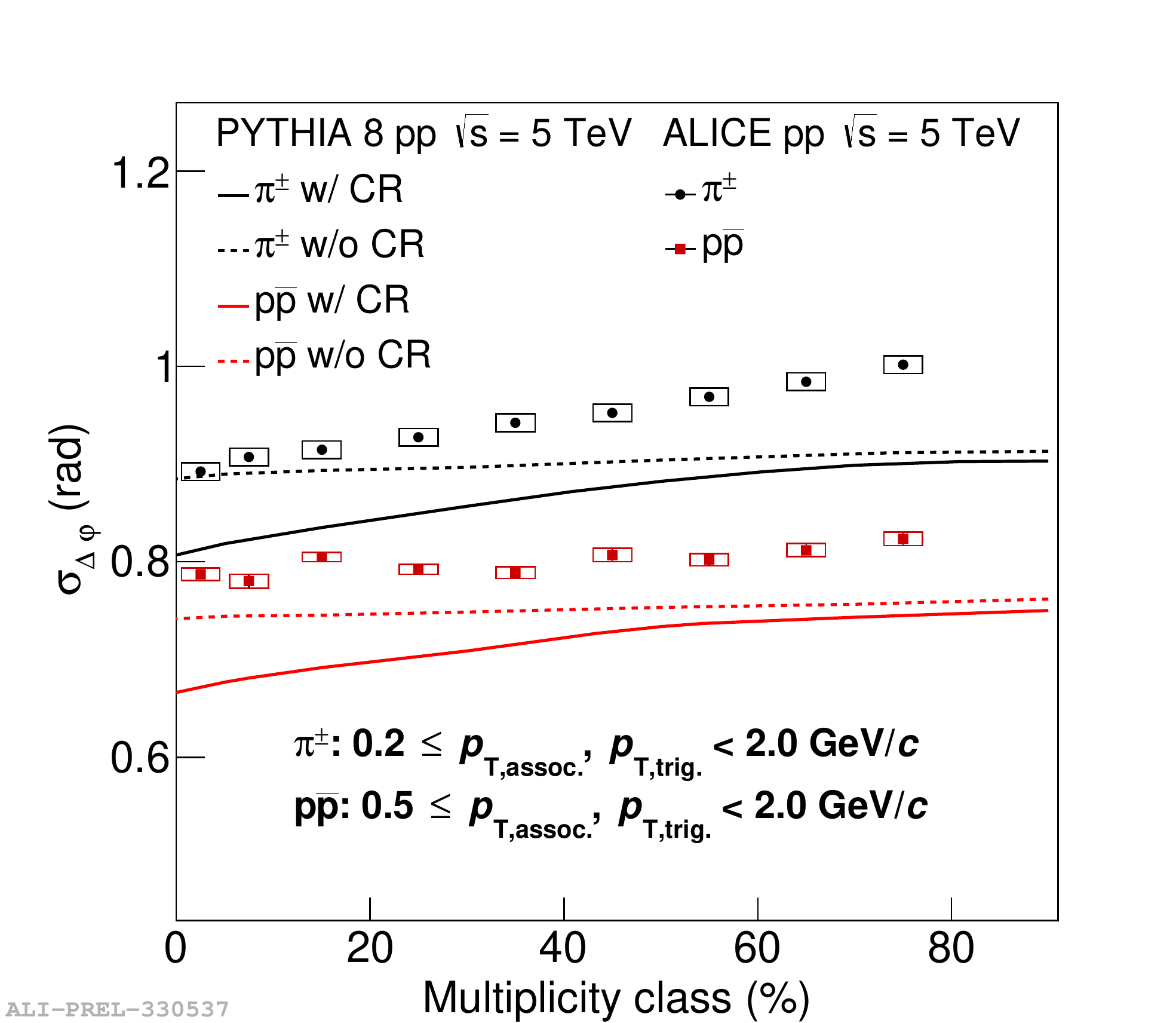}
\caption{}
\end{subfigure}
\begin{subfigure}{0.499\textwidth}
\includegraphics[width=\textwidth]{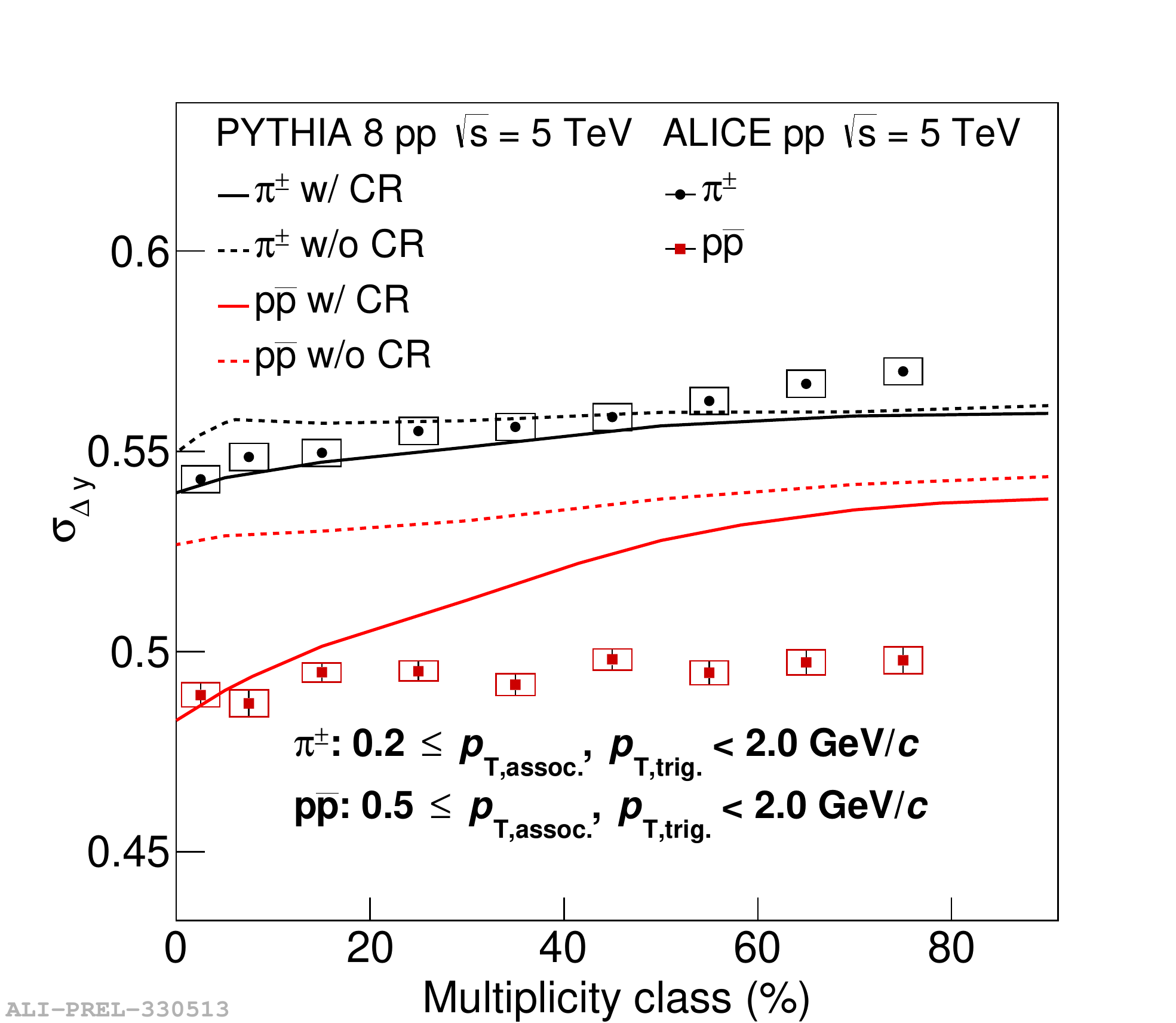}
\caption{}
\end{subfigure}
}
\caption{Balance function widths in \textbf{(a)} $\Delta\varphi$ and \textbf{(b)} $\Delta y$, for low-momentum pions and protons, compared to PYTHIA8 simulations with or without colour reconnection enabled.}
\label{fig:balance_function_sigma}
\end{figure}

In this low-momentum range, it has been observed in both Pb--Pb, p--Pb, and pp collisions that the near-side balance function is narrowing with increasing multiplicity \cite{Balance functions}. In Pb--Pb collisions, this is expected to be due to radial flow of the QGP, but in small systems an alternative explanation is being looked for. One such possibility would be that this is due to colour reconnection, which adds some collective behaviour to the system \cite{Colour reconnection}. For radial flow, there is a mass hierarchy, which is also reflected in the narrowing of the balance function. Therefore, the multiplicity dependence is tested separately for pions and protons and compared to PYTHIA8 with or without colour reconnections, which is shown in Fig. \ref{fig:balance_function_sigma}. For pions, the results agree qualitatively with colour reconnection enabled, but not quantitatively. The proton balance functions show no significant narrowing, which does not agree well with either model. This disfavours colour reconnections as the source of balance function narrowing.

\subsection{$\it \Xi-$hadron correlations at 13 TeV}
\begin{figure}[htb]
\centerline{
\begin{subfigure}{0.333\textwidth}
\includegraphics[width=\textwidth]{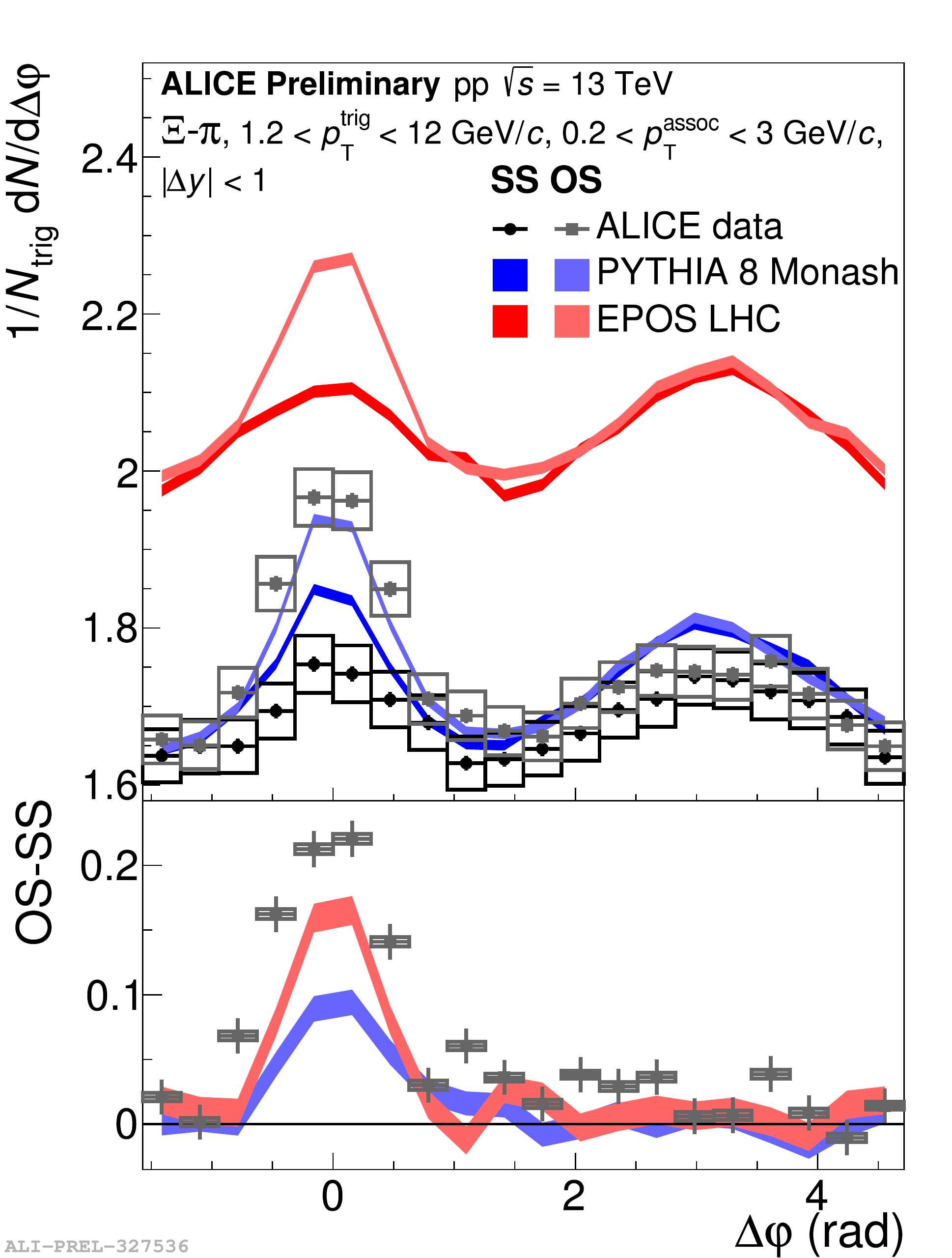}
\caption{}
\end{subfigure}
\begin{subfigure}{0.333\textwidth}
\includegraphics[width=\textwidth]{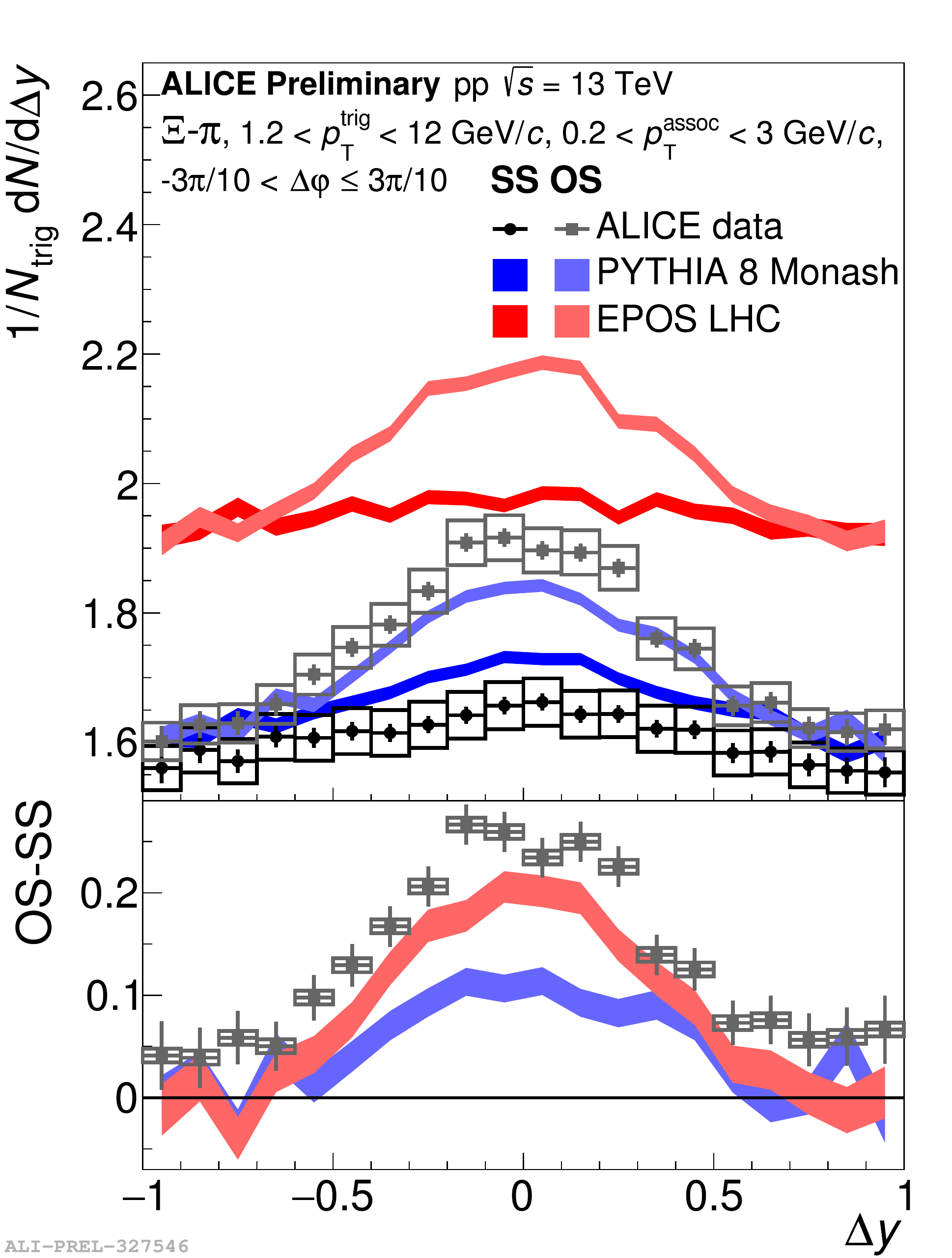}
\caption{}
\end{subfigure}
\begin{subfigure}{0.333\textwidth}
\includegraphics[width=\textwidth]{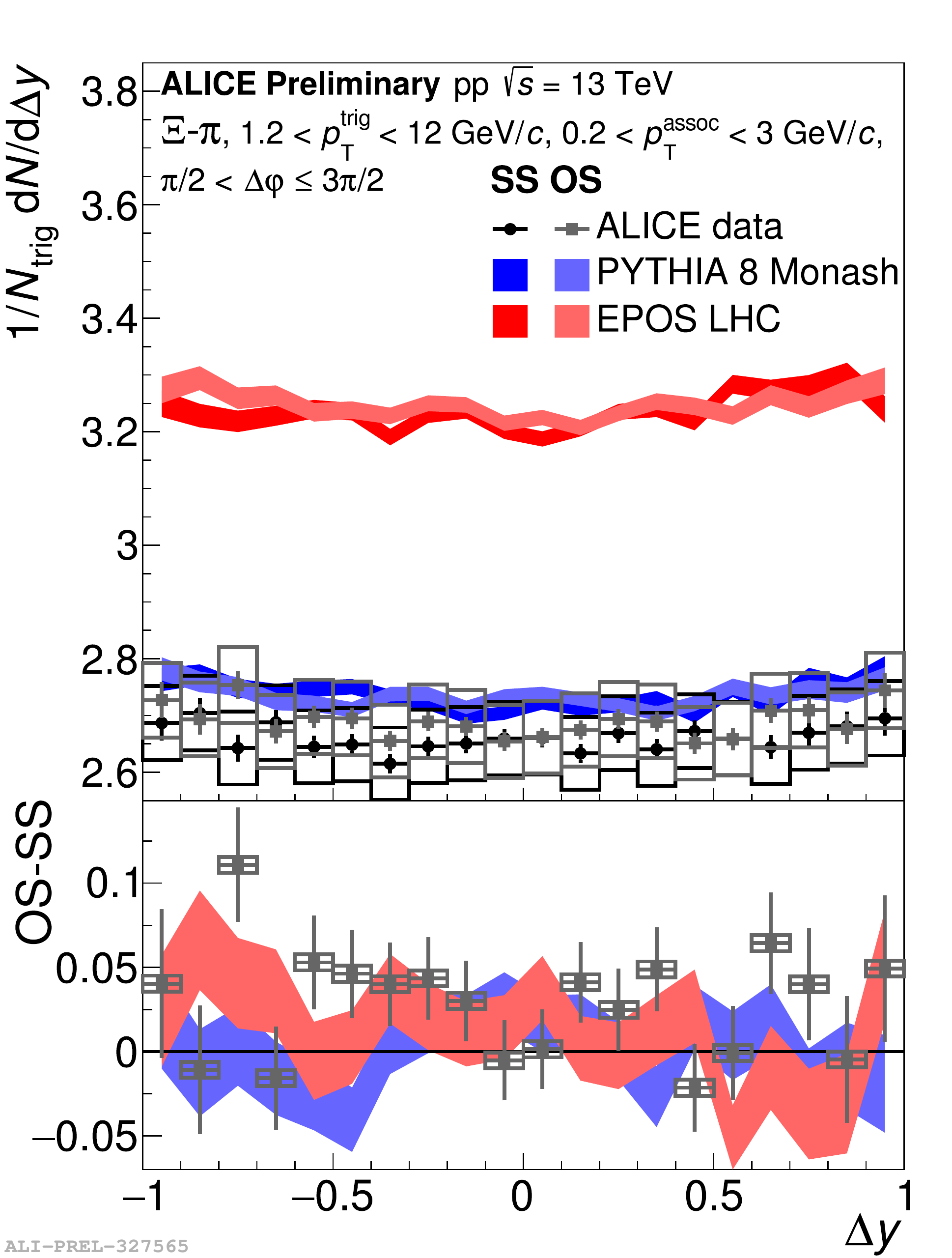}
\caption{}
\end{subfigure}
}
\caption{Per-trigger yields of $\rm\Xi-\pi$ correlations compared with predictions from PYTHIA8 and EPOS-LHC, projected onto \textbf{(a)} $\Delta\varphi$, \textbf{(b)} $\Delta y$ on the near side, and \textbf{(c)} $\Delta y$ on the away side. The lower panels show the difference between opposite- and same-sign correlations (balance functions).}
\label{fig:Xi-pi correlations}
\end{figure}

\begin{figure}[htb]
\centerline{
\begin{subfigure}{0.333\textwidth}
\includegraphics[width=\textwidth]{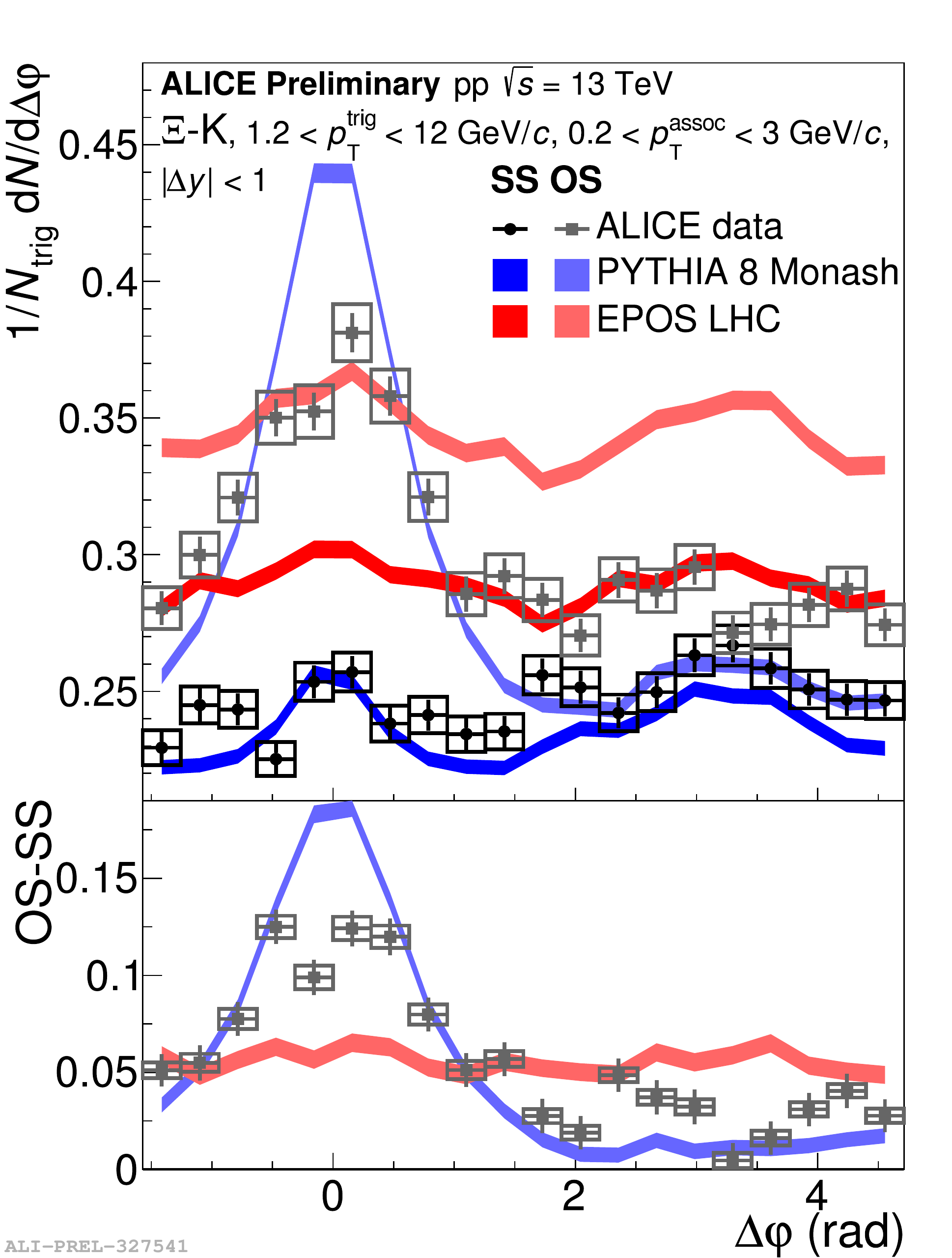}
\caption{}
\end{subfigure}
\begin{subfigure}{0.333\textwidth}
\includegraphics[width=\textwidth]{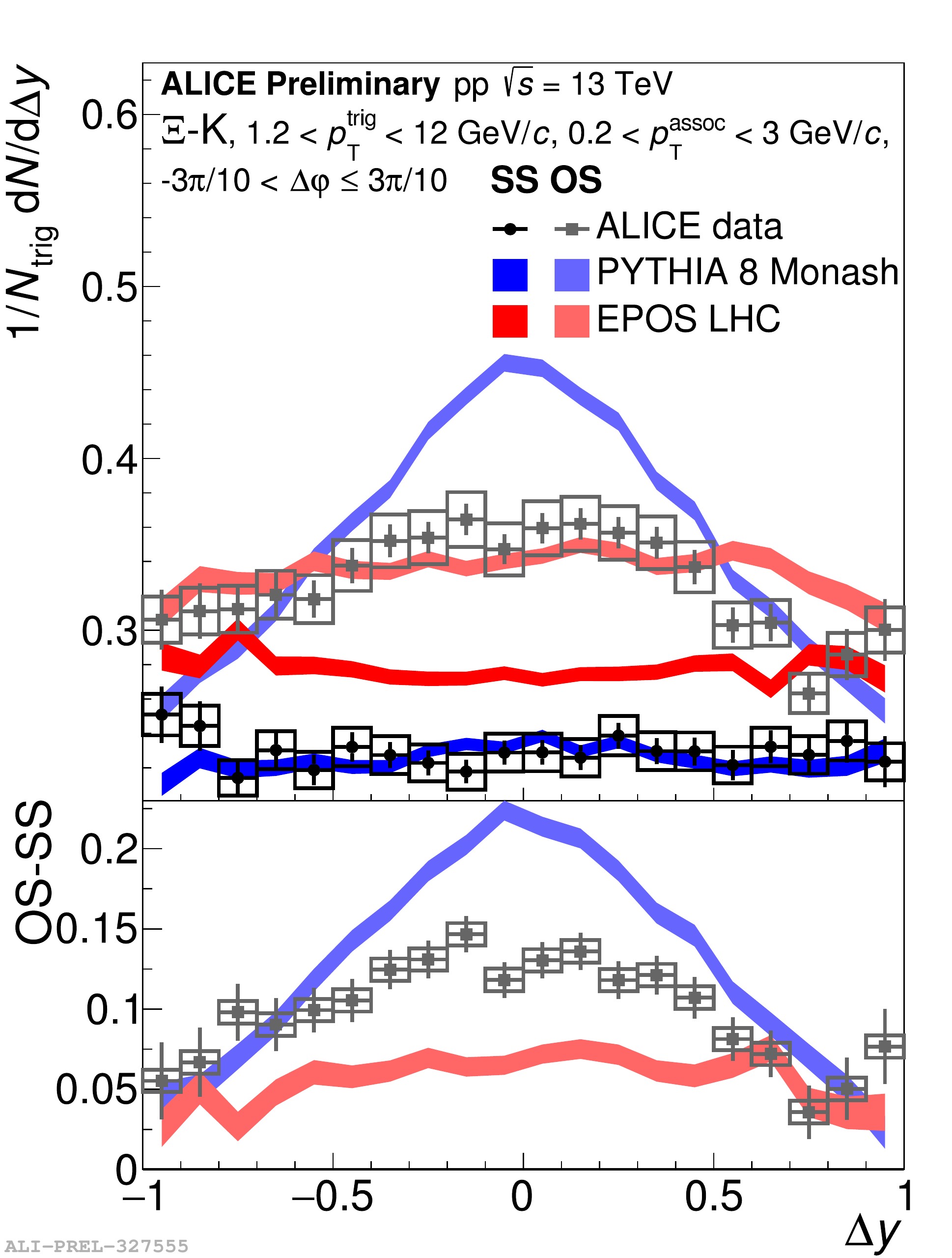}
\caption{}
\end{subfigure}
'\begin{subfigure}{0.333\textwidth}
\includegraphics[width=\textwidth]{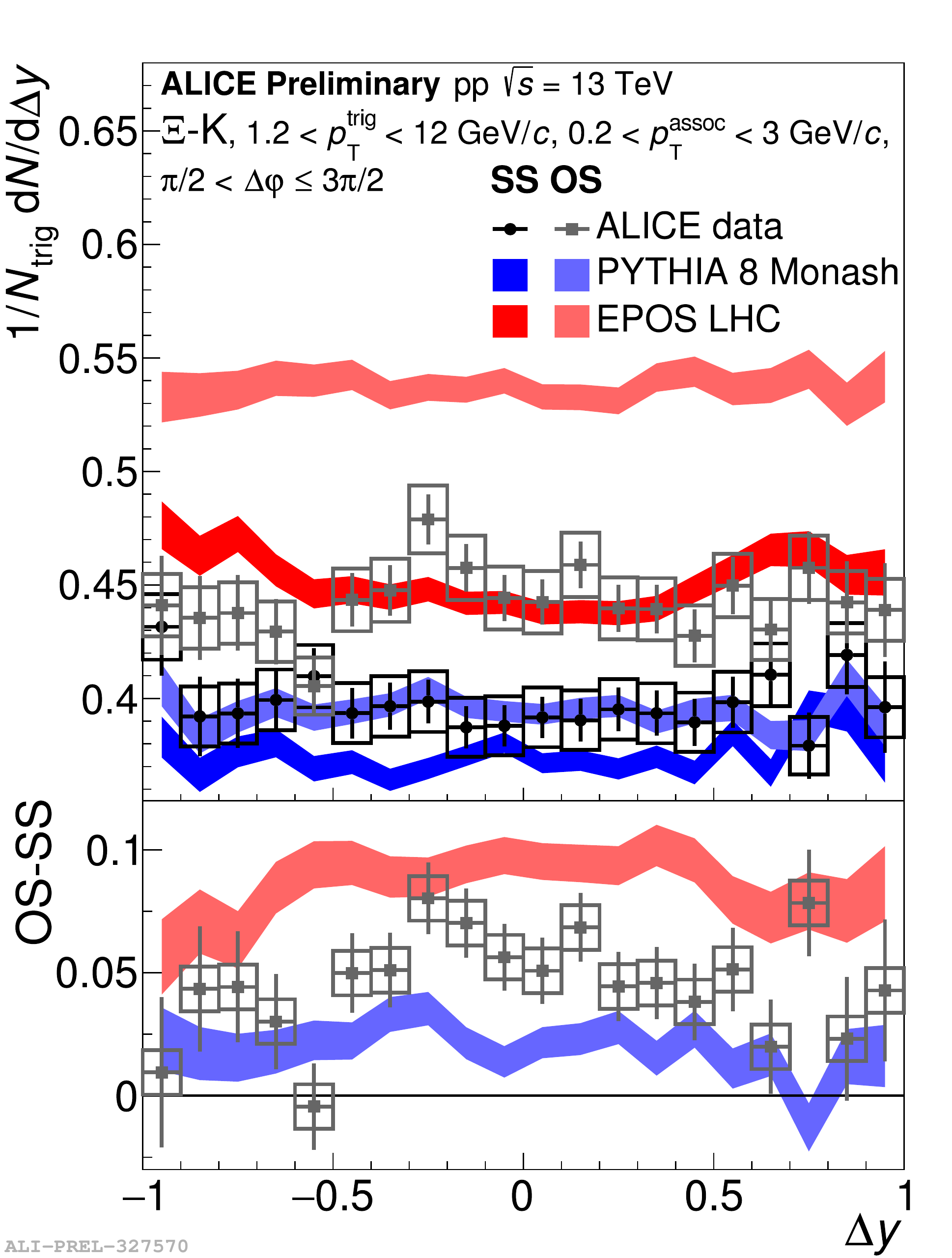}
\caption{}
\end{subfigure}
}
\caption{Figure analogous to Fig. \ref{fig:Xi-pi correlations}, but for $\rm\Xi-K$ correlations.}
\label{fig:Xi-K correlations}
\end{figure}

Per-trigger yields (Eq. \ref{eq:per-trigger_yields}) and balance functions for $\rm \Xi-\pi$ and $\rm \Xi-K$ correlations are shown in Figs. \ref{fig:Xi-pi correlations}-\ref{fig:Xi-K correlations}, along with predictions from PYTHIA8 and EPOS-LHC. This is done to test the strangeness production mechanisms of these models. For $\rm \Xi-\pi$ correlations, both models describe the data quite well qualitatively, although the absolute yields are off by $\sim 20\%$ for EPOS and the balance function by $\sim 50\%$ for PYTHIA. These correlations are likely dominated by minijet fragmentation, which is well taken into account by both models, and will therefore not give much information about strangeness production. For $\rm \Xi-K$ production on the other hand, the near-side peak is stronger than for pions, but wider. For PYTHIA, the near-side peak is much stronger and narrower than in data, whereas for EPOS, the near-side peak is very weak, showing almost no correlation. Therefore, neither a pure string model nor this implementation of a core-corona model are favoured by this measurement. The widening of the near-side peak compared to PYTHIA indicates some collective behaviour, but not as strong as in EPOS-LHC. Also on the away side, there is a significant difference between same- and opposite-sign correlations. Quantitatively, this is somewhere between the two models.

\section{Conclusions}
Angular correlations have proven to be a valuable tool for testing model predictions for several stages of the reaction, including hadron formation. Meson-meson correlations, which are dominated by minijet fragmentation, are well described by QCD inspired models, and particularly PYTHIA8. The same conclusion can be drawn from $\rm\Xi-\pi$ correlations, but given the quantitative agreement, a core-corona model cannot be excluded either.

On the other hand, correlations of baryon pairs with the same baryon number are suppressed on the near side, which is not reproduced by any model. Moreover, $\rm\Xi-K$ correlations are significantly less correlated than in PYTHIA, but not nearly as little as in EPOS, indicating some quark diffusion before the hadronisation phase. For the balance functions, a narrowing with multiplicity is observed at low $p_\mathrm{T}$ for pions, but not for protons. These results challenge models of particle production and more theoretical work is clearly required to understand this.


\end{document}